# ВЕСТНИК ТВЕРСКОГО ГОСУДАРСТВЕННОГО ТЕХНИЧЕСКОГО УНИВЕРСИТЕТА

## Серия «Технические науки»

**Научный журнал**
**№ 1 (1)**

# ОСОБЕННОСТИ ОРГАНИЗАЦИИ ХРАНИЛИЩА ДАННЫХ НА ОСНОВЕ ИНТЕЛЛЕКТУАЛИЗАЦИИ ПОИСКОВОГО АГЕНТА И ЭВОЛЮЦИОННОЙ МОДЕЛИ ОТБОРА ЦЕЛЕВОЙ ИНФОРМАЦИИ

В.К. ИВАНОВ, канд. техн. наук

Тверской государственный технический университет, 170026, Тверь, наб. Аф. Никитина, 22, e-mail: mtivk@mail.ru

В статье представлен систематизированный обзор результатов разработки теоретической основы и пилотной реализации технологии хранилища данных с автоматическим пополнением данными из источников, относящихся к различным тематическим сегментам. Предполагается, что хранилище будет содержать информацию об объектах, обладающих значительным инновационным потенциалом. Механизм селекции такой информации основан на определении ее семантической релевантности генерируемым поисковым запросам. При этом дается количественная оценка инновационности объектов, в частности их технологической новизны и востребованности. Приведено описание принятых показателей инновационности, рассмотрены вопросы применения теории свидетельств для обработки неполной и нечеткой информации. Определены основные идеи методики обработки результатов измерений для расчета вероятностных значений компонентов инновационности, кратко описано применение эволюционного подхода при формировании лингвистической модели архетипа объекта. Приведены сведения об экспериментальной проверке адекватности разработанной вычислительной модели. Результаты исследований, описанные в статье, могут быть использованы для бизнес-планирования, прогнозирования технологического развития, информационного обеспечения экспертизы инвестиционных проектов.

*Ключевые слова:* хранилище данных, интеллектуальный агент, тематический поиск, генетический алгоритм, база данных, свидетельство, бизнес-планирование, инновационность, новизна, востребованность.

## ВВЕДЕНИЕ

В статье впервые представлен систематизированный обзор результатов работ, выполняемых в рамках проекта «Организация и поддержка хранилища данных на основе интеллектуализации поискового агента и эволюционной модели отбора целевой информации». Целью проекта является разработка теоретической основы и пилотной реализации технологии хранилища данных с автоматическим пополнением данными из источников, относящихся к различным тематическим сегментам. Предполагается, что хранилище будет содержать технические, экономические, социальные и другие характеристики объектов, обладающих значительным инновационным потенциалом.

Механизм селекции информации об инновационных объектах основан на определении семантической релевантности такой информации генерируемым поисковым запросам. При этом алгоритмы определения релевантности должны в значительной степени учитывать инновационную составляющую. Основное внимание в проведенных исследовательских работах было уделено разработке соответствующей вычислительной модели. Предлагаемые решения дают количественную оценку инновационности объектов, что в определенной степени является новым подходом.

Актуальность оценки инновационности продуктов очевидна. С экономической точки зрения на микроэкономическом уровне инновации являются ключевым фактором конкурентоспособности любого современного предприятия, как на конкурентных, так и на монопольных рынках. На макроэкономическом уровне активность диффузии инноваций определяет темпы экономического роста, а имплементируемость инноваций способствует концентрации капитала в самых коммерчески эффективных видах предпринимательской деятельности.

Предполагается, что результаты проекта дадут возможность решения следующих прикладных задач:

определение характеристик новых областей и направлений при бизнес-планировании;

прогнозирование технологического развития предприятия, оценка возможностей производства высокотехнологичной продукции;

информационное обеспечение работы экспертных советов, групп, отдельных экспертов.

## ИССЛЕДОВАТЕЛЬСКИЕ ЗАДАЧИ ПРОЕКТА

Основной акцент в исследовательских работах делался на разработке такой вычислительной модели эволюционного процесса генерации запросов и фильтрации результатов поиска, которая будет:

учитывать критерий инновационности объектов, информацию о которых планируется разместить в создаваемом хранилище;

давать количественную оценку релевантности информации о таких объектах;

использовать соответствующие улучшения поискового паттерна.

На первом этапе были сформулированы следующие задачи:

1. Моделирование эволюционного процесса генерации запросов и фильтрации результатов поиска, определение возможностей и условий применения улучшений поискового паттерна.

2. Исследование эффективности предлагаемых решений на тестовых коллекциях (базах данных): экспериментальная проверка адекватности разработанных моделей, обоснование, уточнение и формализация критериев эффективности поискового агента.

## ПРЕДВАРИТЕЛЬНЫЕ РЕЗУЛЬТАТЫ

1. Сформулирован понятийный базис и предложена лингвистическая модель архетипа искомого объекта, образующая поисковый паттерн. Предложены формальные выражения для вычисления значений индикаторов инновационности объектов хранилища: технологической новизны и востребованности потребителями. Введено понятие архетипа объекта, обладающего свойствами, значимыми и общими для всех представителей этого архетипа. Введение понятия архетипа объекта позволяет выполнить формализацию его лингвистической модели, вводя ограничения на эталонную информационную модель в заданной предметной области.

2. Введены нечеткие показатели вероятности того, что объект обладает заданными свойствами инновационности. Для вычисления соответствующих показателей обосновывается применение теории Демпстера – Шафера. Подготовлен детальный обзор публикаций по современным направлениям развития теории Демпстера – Шафера и ее приложений. Проведены исследования применимости теории свидетельств для организации взаимодействия интеллектуальных поисковых агентов в гетерогенной информационно-поисковой системе.

3. Разработан начальный вариант методики обработки результатов измерений исходных показателей для расчета вероятностных значений компонентов инновационности объектов.

4. Обоснован и применен эволюционный подход к решению задачи формирования лингвистической модели архетипа объекта и получения эффективного мультимножества поисковых запросов.

5. Предложены варианты методики, предусматривающие обработку результатов измерений, полученных из нескольких источников с представительными базами данных с учетом надежности источника. Разработан алгоритм групповой обработки результатов измерений уровня инновационности объекта.

6. Выполнена экспериментальная проверка адекватности разработанной вычислительной модели. Проведенные эксперименты подтвердили обоснованность применения методов теории свидетельств для обработки результатов измерений показателей инновационности.

7. Проведен анализ показателей инновационности за двадцатилетний период для некоторых объектов. Результаты подтверждают гипотезу о динамике изменений новизны и востребованности архетипа объекта. Подтверждены известные циклические закономерности в соотношениях инноваций и экономического роста.

Результаты исследований неоднократно докладывались на международных и российских научных конференциях, таких как INFORINO-2018, ПТІ'18, КИИ-2018 и др. Список работ, опубликованных по результатам проекта, можно посмотреть по адресу http://elib.tstu.tver.ru/MegaPro/GetDoc/WebDocs/45. Ниже некоторые важные аспекты выполненных исследований раскрываются более подробно.

## ПОКАЗАТЕЛИ ИННОВАЦИОННОСТИ ОБЪЕКТОВ

Нами введены понятия технологической новизны, востребованности и имплементируемости как составных частей критерия инновационности искомого объекта. Под *технологической новизной* понимаются значительные улучшения, новый способ использования или предоставления объекта (субъектами новизны являются потенциальные пользователи или сам производитель). *Востребованность* – это осознанная потенциальным производителем необходимость для него объекта, оформленная в спрос. *Имплементируемость* определяет технологическую обоснованность, физическую осуществимость и способность объекта быть интегрированным в систему для получения желаемых эффектов.

Предложена также лингвистическая модель архетипа искомого объекта, образующая поисковый паттерн. Термы модели классифицируется как ключевые свойства, описывающие структуру объекта, условия применения или результаты функционирования. Область определения архетипа отмечается маркером. Поисковые запросы конструируются как комбинации термов и маркера; при значительном количестве термов используется генетический алгоритм генерации запросов и фильтрации результатов – оригинальная разработка участников проекта, позволяющая получить квазиоптимальный набор поисковых запросов. Предложены следующие формальные выражения для вычисления значений индикаторов инновационности:

$$Nov = 1 - \frac{1}{S}\sum_{k=1}^{S}\left[1 - \exp(1 - \frac{R_k}{R_0^{01}})\right],$$

где *Nov* – новизна объекта; $R_k$ – число документов, найденных в результате выполнения *k*-го запроса в базе данных, содержащей информацию о рассматриваемой предметной области; $R_0^{01}$ – нормированное на диапазон [0; 1] число документов, найденных в результате выполнения запроса, состоящего из одного терма-маркера;

$$Rel = \frac{1}{S}\sum_{k=1}^{S}\left[1-\exp(1-\frac{F_k}{F_0^{01}})\right],$$

где *Rel* – востребованность объекта; $F_k$ – частота выполнения пользователями запросов, аналогичных *k*-му запросу; $F_0^{01}$ – нормированное на диапазон [0; 1] значение частоты выполнения пользователями запроса, состоящего из одного терма-маркера.

Используется гипотеза об адекватности отображения в информационном пространстве реальных процессов при условии достаточного количества свидетельств о них из разных источников. Оценка новизны объекта основывается на нормализованной интегральной оценке количества результатов поиска информации об объекте в гетерогенных базах данных. Предполагается, что для новых объектов количество результатов поиска, релевантных поисковому паттерну, будет меньше, чем для давно существующих и известных объектов. Оценка востребованности объекта основывается на нормализованной интегральной оценке частоты выполнения пользователями запросов, аналогичных запросам, генерированным из поискового паттерна. Оригинальность решения заключается в согласованном получении и использовании информации об объектах хранилища из многих гетерогенных источников, включая информационные интернет-ресурсы. Учитывая непосредственную количественную оценку инновационности, мы считаем этот подход дополняющим традиционные [1, 2].

### ПРИМЕНЕНИЕ ТЕОРИИ СВИДЕТЕЛЬСТВ

В связи с тем, что ожидается очевидная неполнота и неточность информации об объектах, полученной из различных источников, в проекте вводятся нечеткие показатели вероятности того, что объект обладает технологической новизной, и вероятности того, что объект востребован. Для вычисления указанных вероятностей обосновано применение теории свидетельств Демпстера – Шафера [3, 4]. Так, принимается, что базовая вероятность *m* попадания результатов измерения показателя инновационности объекта (*Nov* или *Rel*) в интервал значений *A* определяется следующим образом:

$$m: P(\Omega) \to [0;1], m(\varnothing) = 0, \sum_{A \in P(\Omega)} m(A) = 1,$$

где $\Omega$ – множество значений результатов измерения показателя; $P(\Omega)$ – множество всех подмножеств $\Omega$. Для заданных *k* интервалов рассчитываются функция доверия $Bel(A) = \sum_{A_k: A_k \subseteq A} m(A_k)$ и функция правдоподобия $Pl(A) = \sum_{A_k: A_k \cap A} m(A_k)$, которые определяют верхнюю и нижнюю границы вероятности обладания объектом заданным свойством. Таким образом дается оценка значений показателей *Nov* или *Rel* в условиях неполноты и неточности информации об объектах. Соответствующие специализированные алгоритмы являются новой областью применения теории свидетельств.

В рамках проекта были проведены исследования применимости теории свидетельств для решения задач диагностики сложных технических систем и оптимального управления эволюцией многостадийных процессов в нечеткой динамической среде [5]. Цель этих исследований – предложить новую архитектуру для взаимодействия интеллектуальных поисковых агентов в гетерогенной информационно-поисковой системе, основанную на понятии «аномальности» агента. Аномальное состояние (АС) поискового агента интерпретируется как наличие претендента на инновацию в результатах поиска этого агента и диагностируется как выход индикаторов инновационности объекта за характеристические значения.

Для индикации АС поискового агента $s \in S$ используется следующее необходимое и достаточное условие: $(\forall s \in S)(F = 0) \leftrightarrow P^* \neq 0$, где $F$ – индикаторная функция; $P^*$ – множество зарегистрированных АС. Учитывая потенциально большое количество источников информации, при проверке диагностических гипотез необходимо не пропустить действительную инновацию и избежать индикации ложной инновации. Индикаторная функция дает нам возможность количественного определения «рациональности» и «полезности» поискового агента в заданном контексте.

## МЕТОДИКА РАСЧЕТА ИННОВАЦИОННОСТИ ОБЪЕКТОВ

Разработан начальный вариант методики обработки результатов измерений исходных показателей для расчета вероятностных значений компонентов инновационности объектов: числа релевантных документов и частот выполнения запросов пользователями релевантных запросов.

Основные шаги методики:

1. Выполнение заданного количества квазиоптимальных поисковых запросов, сгенерированных с помощью генетического алгоритма из поискового паттерна. С точки зрения теории свидетельств эти запросы являются наблюдаемыми подмножествами или фокальными элементами.

2. Определение базовой вероятности появления признаков инновационности для каждого из наблюдаемых подмножеств.

3. Вычисление функций доверия и правдоподобия для заданного количества интервалов группирования, соответствующих номинальной шкале полного или частичного наличия / отсутствия признаков инновационности.

Варианты методики предусматривают обработку результатов измерений, полученных из нескольких источников (поисковых систем) с учетом коэффициента дисконтирования для базовой вероятности. В качестве источников результатов измерений выступают гетерогенные поисковые системы, имеющие представительные базы данных. Использовалась классическая модель комбинирования различных свидетельств, основанная на правиле Демпстера. По имеющимся результатам измерений рассчитывается коэффициент конфликтности, который используется при вычислении комбинированной базовой вероятности появления признаков инновационности для искомого объекта. Методика не имеет аналогов и обладает потенциалом для применения других моделей комбинирования свидетельств.

Предложен также собственный алгоритм групповой обработки результатов измерений уровня инновационности объекта. Алгоритм предусматривает операции сбора исходных данных, ввод данных по категориям, расчет базовых вероятностей свидетельств, комбинирование свидетельств по компонентам объекта и показателям, определение значения функций доверия и правдоподобия комбинированных свидетельств, формирование интегральных оценок по компонентам объекта и / или показателям.

Комбинирование выполняется рекурсивно по парам источников: из двух источников свидетельств образуется один условный источник, свидетельства которого комбинируются со следующим фактическим источником. Разработанный алгоритм реализован в виде веб-приложения с графическим интерфейсом пользователя. Мы предполагаем внедрение этого приложения в различные сферы инженерной деятельности, где возможно применение апостериорных источников информации, таких как электронные датчики или цифровые измерительные устройства. Другим направлением развития является поддержка математических методов, альтернативных теории свидетельств.

## ЭВОЛЮЦИОННЫЙ ПОДХОД К ФОРМИРОВАНИЮ ЛИНГВИСТИЧЕСКОЙ МОДЕЛИ ОБЪЕКТА

В проекте обоснован и применен эволюционный подход к решению задачи формирования лингвистической модели архетипа объекта. Основная идея состоит в том, чтобы организовать с помощью специального генетического алгоритма (ГА) эволюционный процесс, генерирующий стабильную и эффективную совокупность поисковых запросов для получения наиболее релевантных результатов. Начальная популяция из поисковых запросов представлена в виде множества векторов, каждый из которых есть набор ключевых терминов. При протекании процесса закодированные запросы последовательно подвергаются генетическим изменениям (скрещиванию и мутациям) и выполняются в поисковой системе. Затем оценивается семантическая релевантность промежуточных результатов поиска, вычисляются значения фитнес-функции и выбираются наиболее подходящие запросы. В результате эволюционного процесса получаем мультимножество эффективных запросов, которое интерпретируется как лингвистическая модель целевого объекта. Оригинальной особенностью предлагаемого эволюционного подхода, подробно описанного в [6], является способ вычисления значений фитнес-функции и интерпретации генетических операций.

В понимании работы ГА важную роль играет фундаментальная теорема схем Холланда. Она сформулирована применительно к каноническому ГА и доказывает его способность генерировать достаточное количество эффективных схем особей для достижения окрестностей оптимума фитнес-функции за конечное число шагов. Очевидна целесообразность проверки на выполнение теоремы Холланда любых модификаций канонического ГА, созданных для решения конкретной задачи. В наших исследованиях определены условия корректной проверки выполнения теоремы Холланда, в частности требования к кодированию генотипа. Предложен новый метод кодирования генотипа, который использует расстояние между векторами. Выявлены актуальные направления доработки программного обеспечения ГА, а именно создание алгоритмов расчета весовых коэффициентов для компонентов фитнес-функции и техника синонимизации ключевых терминов.

В части проведения опытно-конструкторских работ было выполнено тестирование программного интерфейса для массовой индексации документов различных форматов, размещаемых в хранилище, в среде серверного программного обеспечения Apache Lucene Solr (http://lucene.apache.org), а также DocFethcher (локальной имплементации индексатора Lucene).

## ЭКСПЕРИМЕНТАЛЬНЫЕ ИССЛЕДОВАНИЯ

Для экспериментальной проверки адекватности разработанной вычислительной модели были сформулированы следующие задачи:

1. Апробация методики расчета показателей инновационности объектов в соответствии с принятой вычислительной моделью.

2. Сравнение рассчитанных результатов показателей инновационности с их экспертными оценками.

3. Сравнение рассчитанных результатов показателей инновационности, полученных после обработки данных из разных поисковых систем.

4. Оценка динамики изменений показателей инновационности объекта во времени.

5. Получение обоснования пригодности измеренных значений индикаторов инновационности для дальнейшей обработки.

В качестве источников информации об объектах были отобраны следующие базы данных: http://new.fips.ru, https://elibrary.ru, https://rosrid.ru, https://yandex.ru, https://wordstat.yandex.ru, https://google.com, https://adwords.google.com, https://patents.google.com, https://scholar.google.ru. Объектами для проведения анализа стали десять лучших изобретений 2017 года, выбранные экспертами Роспатента, и десять случайно выбранных изобретений, зарегистрированных в том же году. Поисковые паттерны готовились экспертами. Корпуса документов для анализа формировались в поисковых системах указанных выше баз данных.

Проведенные эксперименты подтвердили обоснованность применения методов теории свидетельств для обработки результатов измерений показателей инновационности. Несмотря на то, что абсолютные значения измерений ожидаемо различаются для разных источников данных, разработанная модель адекватно оценивает относительные изменения значений показателей новизны и востребованности объекта (для комбинированных значений показателей прогнозируются аналогичные результаты). Средние значения новизны патентов, лучших по оценке экспертов, превышают средние значения новизны случайно выбранных патентов. Аналогично, средние значения востребованности лучших объектов превышают средние значения востребованности случайно выбранных объектов. Таким образом, вычислительная модель подтверждает экспертную оценку инновационности.

В ходе проведения экспериментов был выполнен анализ новизны объектов, оцененной за определенный (двадцатилетний) период. На рис. 1 и 2 в качестве примера представлены графики изменения новизны и востребованности архетипа потенциально инновационного объекта «Электронный индикатор уровня». Аппроксимация полученных значений (сплошные линии тренда на графиках) подтверждает гипотезу об уменьшении новизны объекта во времени. Значение востребованности объекта со временем возрастает. Очевидно, что за время своего существования объект приобретает все большую популярность среди пользователей, поэтому потенциальный интерес к нему растет. Следует отметить, что значения показателей инновационности во времени фактически рассчитывались не для какого-либо конкретного объекта (например, патента), а для его архетипа, заданного лингвистической моделью патента.

В результатах экспериментов нашли подтверждение известные циклические закономерности, выявленные при анализе соотношений инноваций и экономического роста. Динамика показателей инновационности на достаточно большом интервале имеет циклический характер (пунктирные линии тренда на графиках). Более того, мы видим корреляцию между значениями новизны и востребованности архетипа объекта. Хотя результаты вычислений в целом неоднозначны, мы идентифицируем цикличность, которая требует проверки гипотезы инновационных циклов в конкретной области применения.

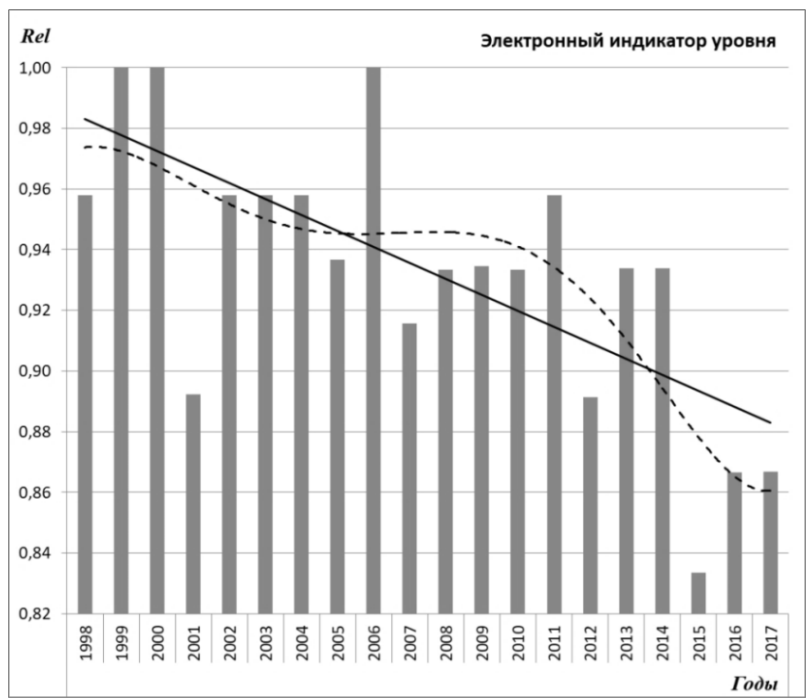

Рис. 1. Изменение показателя инновационности объекта
«Технологическая новизна» (пример)

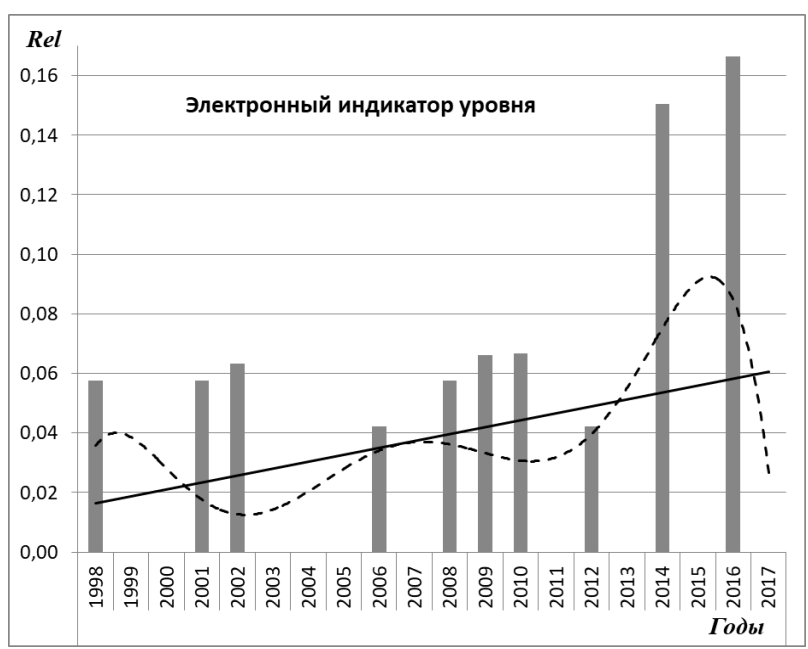

Рис. 2. Изменение показателя инновационности объекта
«Востребованность» (пример)

**ЗАКЛЮЧЕНИЕ**

Очевидно, что цели обсуждаемого в статье проекта соответствуют приоритетному направлению развития критических технологий информационных, управляющих, навигационных систем. Текущие и планируемые результаты, по нашему мнению, могут способствовать переходу к передовым цифровым, интеллектуальным производственным технологиям, роботизированным системам, созданию систем обработки больших объемов данных.

На следующем этапе выполнения проекта планируется получить:

формальное описание лингвистической модели архетипа объекта, учитывающее особенности поискового паттерна, ориентированного на инновационность искомых объектов;

результаты экспериментов, подтверждающих гипотезу о применимости поискового паттерна для объектов, имеющих инновационный потенциал;

архитектуру и модель поведения поискового интеллектуального агента, работающего с источником данных в системе гетерогенных баз данных;

прототип программного обеспечения многоагентной поисковой системы, обеспечивающей совместную обработку результатов измерений показателей инновационности искомых объектов.

# PECULIARITIES OF ORGANIZATION OF DATA STORAGE BASED ON INTELLIGENT SEARCH AGENT AND EVOLUTIONARY MODEL SELECTION THE TARGET INFORMATION

V.K. IVANOV, Cand Sci

Tver State Technical University, 22, Af. Nikitin emb., 170026, Tver, Russian Federation, e-mail: mtivk@mail.ru

The article presents a systematic review of the results of the development of the theoretical basis and the pilot implementation of data storage technology with automatic replenishment of data from sources belonging to different thematic segments. It is expected that the repository will contain information about objects with significant innovative potential. The mechanism of selection of such information is based on the determination of its semantic relevance to the generated search queries. At the same time, a quantitative assessment of the innovation of objects, in particular their technological novelty and demand is given. The article describes the accepted indicators of innovation, discusses the application of the theory of evidence for the processing of incomplete and fuzzy information, identifies the main ideas of the method of processing the results of measurements for the calculation of the probabilistic value of the components of innovation, briefly describes the application of the evolutionary approach in the formation of the linguistic model of the archetype of the object, provides information about the experimental verification of the adequacy of the developed computational model. The research results that are described in the article can be used for business planning, forecasting of technological development, information support of investment projects expertise.

*Keywords:* data warehouse, intelligent agent, thematic search, genetic algorithm, database, evidence, business planning, innovation, novelty, demand.

## ACKNOWLEDGMENTS

The author expresses deep gratitude to colleagues – project participants who help to prepare this article: N.V. Vinogradova, A.G. Glebova, A.A. Gusarov, E.N. Kabanova, P.I. Meskin, I.V. Obraztsov, B.V. Paluh, A.N. Sotnikov.

This work was financially supported by the Russian Foundation for Basic Research, project № 18-07-00358 A.